\documentclass[preprint,aps,prb]{revtex4-1}

\usepackage{graphicx}
\usepackage{amsmath}
\usepackage{mathtools}

\def\BFA{{Bijl-Feynman approximation}}
\def\qW{{\bf W}}                                   
\def\qT{{\bf T}}                                   
\def\qS{{\bf S}}                                   
\def\qr{{\bf r}}                                   

\begin{document}

\title{Dipolar Bilayer with Antiparallel Polarization -- a Self-Bound Liquid}

\author{Martin Hebenstreit$^{1,2}$, Michael Rader$^{1,2}$, and R. E. Zillich$^2$}

\affiliation{$^1$Institute for Theoretical Physics, University of Innsbruck,
Technikerstr. 25, 6020 Innsbruck, Austria\\
$^2$Institute for Theoretical Physics, Johannes Kepler University, Altenbergerstrasse 69, 4040 Linz, Austria
}

\begin{abstract}
Dipolar bilayers with antiparallel polarization, i.e. opposite polarization
in the two layers, exhibit liquid-like rather than gas-like behavior.  In particular,
even without external pressure a self-bound liquid puddle of constant density
will form.  We investigate the symmetric case of
two identical layers, corresponding to a two-component Bose system with equal
partial densities.
The zero-temperature equation of state $E(\rho)/N$, where $\rho$ is the total density,
has a minimum, with an equilibrium density that decreases with increasing distance between
the layers.  The attraction necessary for a self-bound liquid comes from the
inter-layer dipole-dipole interaction that leads to a mediated intra-layer attraction.
We investigate the regime of negative pressure towards
the spinodal instability, where the bilayer is unstable against
infinitesimal fluctuations of the total density, conformed by calculations of the
speed of sound of total density fluctuations.
\end{abstract}


\maketitle

{\em Introduction.}
Experiments with Bose gases of atoms with large magnetic moments
($^{52}$Cr~\cite{griesmaierPRL05,lahayeNature07},
$^{164}$Dy~\cite{luPRL11dysprosium}, $^{168}$Er,~\cite{aikawaPRL12})
are fueling the interest to understand the effects of the dipole-dipole
interaction (DDI) on stability, shape, and dynamics
of dipolar Bose condensates
(reviewed in Refs.~\cite{baranovPhysRep08,lahayeRepProgPhys09,baranovChemPhys12}).
The strength of the DDI can be characterized by the dipole length $r_D=m D^2/(4\pi\varepsilon_0\hbar^2)$,
where $m$ is the mass of the dipolar atom or molecule, and $D$ is its dipole moment.
The value of $r_D$ can be compared with the average inter-particle spacing, $r_s\sim \rho^{-1/m}$,
where $\rho$ is the number density of the condensate and $m$ the dimensionality.
For $r_D\ll r_s$, the DDI is weak; in general, other contributions to the interaction, such
as the s-wave scattering length $a$, will dominate (except if $a$ is tuned to a
sufficiently small value~\cite{lahayeNature07}).  For $r_D \gtrsim r_s$, the DDI
will be the dominant interaction.
The magnetic DDI is usually negligible, only for the handful of atoms
mentioned above, its effect has been observed, but it is difficult
to increase the density such that $r_D \gtrsim r_s$.
Compared with the magnetic dipole moment of atoms, the electric dipole moment of
heteronuclear molecules can be orders of magnitude larger, leading to large values
for $r_D$ (e.g. $r_D=5\times 10^5\AA$ for a fully polarized NaCs).
Association of two atoms using a Feshbach resonance and transfer to
the rovibrational ground state has been achieved for example for
$^{7}$Li$^{133}$Cs \cite{deiglmayrPRL08},
$^{40}$K$^{87}$Rb \cite{niScience08},
$^{41}$K$^{87}$Rb \cite{aikawaPRL10} and 
$^{85}$Rb$^{133}$Cs \cite{sagePRL05,takekoshiPRA12}.
But it remains a challenge to produce a degenerate quantum gas of dipolar molecules.

The anisotropy of the DDI leads to a measurable anisotropy of the
speed of sound \cite{bismutPRL12}, but also
an anisotropic superfluid response~\cite{ticknorPRL11} has been predicted.
The attractive part of the DDI can give
rise to roton or roton-like excitations in a dipolar Bose gas
layer~\cite{santosPRL03,odellPRL03,hufnaglJLTP10,hufnaglPRL11,hufnaglPRA13}.  An anisotropic 2D quantum gas can
be realized by tilting the polarization dipoles in a deep trap, and
a stripe phase can form spontaneously~\cite{maciaPRA11,maciaPRL12}.
For $r_D \gg r_s$, dipoles will crystallize without
imposing an optical lattice~\cite{astraPRL07,buechlerPRL07,matveevaPRL12}.  A bilayered dipolar Bose gas can
dimerize if the polarization direction in the two layers is the same~\cite{maciaPRA14}.  Also the
case of antiparallel polarization in two layers has been studied, where dipoles
are perpendicular to the layer, but the orientation of the dipoles in one layer is opposite
to that in the other layer~\cite{lechnerPRL13}.

In this work we study such a bilayer of bosonic dipoles with antiparallel polarization.  The key result
is that it is a self-bound liquid, unlike bilayers with parallel polarization, and therefore
does not need external pressure in the form of a trap potential to stay together.
We show that the liquid nature is a consequence of the attractive part of the inter-layer DDI,
which leads to cohesion due to ``dipole bridges'' that effectively act as a glue to
bind all particles together.  For our calculations we use a variational many-body theory,
the hypernetted-chain Euler-Lagrange method (HNC-EL), which includes pair correlations.
For comparison and validation, we use path integral Monte Carlo (PIMC) simulations.

{\em Methodology.}
A 1D optical lattice slices a BEC into quasi-2D layers separated by
a distance $d$.  Since the dipole length $r_D$ can easily exceed the typical $d$ value of about 500nm,
the DDI interaction between dipoles in different layers can lead to appreciable coupling.
We consider here two translationally invariant layers $A$ and $B$, approximate each layer as
two-dimensional, and assume no tunneling occurs.
With these simplifications we get two coupled 2D systems, i.e. a binary Bose mixture.
The particles in the two layers shall be the same molecules, hence with same mass
and dipole moment.  In units of dipole length $r_D$ and the dipole energy $E_D=\hbar^2/(m r_D^2)$,
the Hamiltonian is
$$
H=-{1\over 2}\sum_{\alpha,i} \nabla^2_{i,\alpha}
+{1\over 2}\sum_{\alpha,\beta}\sideset{}{'}\sum_{i,j} v_{\alpha,\beta}(|\qr_{i,\alpha} - \qr_{j,\beta}|)\ .
$$
$\alpha$ and $\beta$ index the layer, $\alpha,\beta\in \{A,B\}$, and $i$ the particles
within a layer.
The primed sum indicates that for $\alpha=\beta$ we only sum over $i\ne j$.
$v_{\alpha,\beta}(|\qr_{i,\alpha} - \qr_{j,\beta}|)$ is the DDI, in units of $E_D$, between
dipole $i$ at $\qr_{i,\alpha}$ in layer $\alpha$ and dipole $j$ at $\qr_{j,\beta}$ in layer $\beta$.
We neglect short-ranged interactions compared to the DDI.
The intralayer interaction ($\alpha=\beta$) is purely repulsive, $v_{\alpha,\alpha}(r)={1/r^3}$.
The interlayer interaction, $\alpha\ne\beta$,
is $v_{AB}(r)=(2d^2-r^2)/(d^2+r^2)^{5/2}$, which is repulsive for small $r$, but attractive
for large $r$, and has a minimum at $r_{\rm min}=2d$.
Since the average interlayer interaction vanishes, $\int d^2r\, v_{AB}(r)=0$,
the coupling between layers in the ground state would vanish in a mean field approximation
and the ground state energy would just be the sum of the energies of each layer.

For the many-body ground state we use the variational Jastrow-Feenberg
ansatz~\cite{FeenbergBook} consisting of a product of pair correlation functions for
a multi-component Bose system,
$
  \Psi_0 = \exp\big[{1\over 4}\sum_{\alpha,\beta}\sum'_{i,j} u_{\alpha,\beta}(|\qr_{i,\alpha} - \qr_{j,\beta}|)\big]
$.
Higher order correlations $u_{\alpha,\beta,\gamma}(\qr_{i,\alpha},\qr_{j,\beta},\qr_{k,\gamma})$ could
be included, as is routinely done for single-component calculations.  Past experience
has shown that triplet correlations improve the ground state energy, leading to results very
close to exact QMC simulations, but they do not change the qualitative picture.
We therefore restrict ourselves to pair correlations, but check the results against PIMC simulations.
The pair correlations $u_{\alpha,\beta}(r)$ are determined from Ritz' variational principle,
i.e.\ from the coupled Euler-Lagrange equations,
$\delta\langle\Psi_0|H|\Psi_0\rangle / \delta u_{\alpha,\beta}(r) = 0$.  They are solved by
expressing $u_{\alpha,\beta}(r)$ in terms of the pair distribution function
$
  g_{\alpha,\beta}(|\qr_{\alpha} - \qr_{\beta}|)
= {N_\alpha(N_\beta-\delta_{\alpha\beta}) \over \rho_\alpha\rho_\beta} \int\! d\tau_{\alpha,\beta}|\Psi_0|^2
$
where the integral is over all particles except one in layer $\alpha$ and one in layer
$\beta$, and $\rho_\alpha=N_\alpha/V$ is the partial density of component $\alpha$.
$u_{\alpha,\beta}(r)$ and $g_{\alpha,\beta}(r)$ are related via the hypernetted-chain equations,
and the equations to be solved for $g_{\alpha,\beta}(r)$ are the hypernetted-chain
Euler-Lagrange (HNC-EL) equations.  Certain ``elementary'' diagrams
cannot be summed up exactly, and have to be approximated.  We simply neglect them completely
(HNC-EL/0 approximation);
what we said about neglecting triplet correlations also applies to neglecting elementary diagrams. 
Details about the HNC-EL method can be found in
Refs.~\cite{kroValencia98,krotscheckPhysRep93}, and particularly for
Bose mixtures in Refs.~\cite{Chuckmix,OldChuckMix,chakrabortyPRB82}.  The HNC-EL/0 equations
for an arbitrary number of components are (bold-faced capital letters denote matrices):
\begin{align*}
\qW(k)  & = -{1\over 2}\Big[\qS(k) \qT(k)+\qT(k) \qS(k) -3\qT(k) \\
        & +\qS^{-1}(k)\qT(k)\qS^{-1}(k)\Big] \\
V^{ph}_{\alpha,\beta}(r)& = g_{\alpha,\beta}(r) v_{\alpha,\beta}(r)
          + {\hbar^2\over 2m_{\alpha,\beta}} \Big|\nabla\sqrt{g_{\alpha,\beta}(r)}\Big|^2 \\
         &+ (g_{\alpha,\beta}(r)-1)W_{\alpha,\beta}(r) \\
{\bf V}^{ph}(k) &= \qS^{-1}(k)\qT(k)\qS^{-1}(k) - \qT(k)
\end{align*}
where $m_{\alpha,\beta}$ is the reduced mass (for the symmetric bilayer,
$\hbar^2/ 2m_\alpha=\hbar^2/ 2m_\beta=1/2$ in dipole units).  $S_{\alpha,\beta}(k)$ is the
static structure function
$S_{\alpha,\beta}(k)=\delta_{\alpha\beta}+\sqrt{\rho_\alpha\rho_\beta}\, {\rm FT}[g_{\alpha,\beta}-1]$
(FT denotes Fourier transformation).
The kinetic energy matrix, $T_{\alpha,\beta}(k)=\delta_{\alpha\beta}(\hbar^2k^2/ 4m_{\alpha,\beta})$,
becomes $T_{\alpha,\beta}(k)=\delta_{\alpha\beta}{k^2\over 2}$ in our case.  The HNC-EL/0 equations
can be solved iteratively.  Usually the convergence is stable and fast, but close to
an instability like the spinodal point discussed below, we use linear mixing between iterations
to ensure convergence.

\begin{figure}[ht]
\begin{center}
\includegraphics*[width=0.64\textwidth]{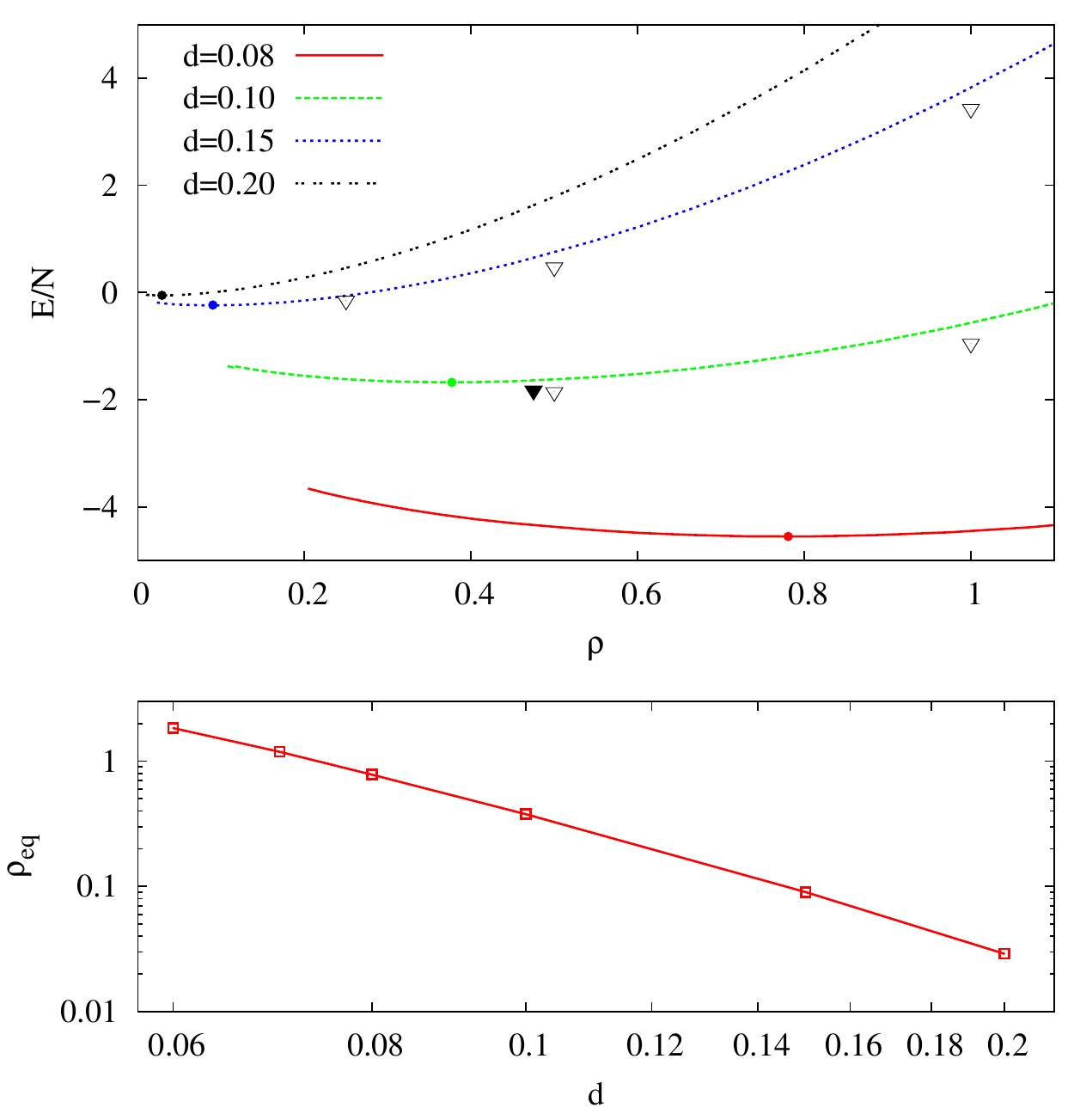}
\end{center}
\caption{
(Color online)
Top panel: Ground state energy per particle $E/N$ versus the total density $\rho$,
for several layer distances $d$. 
The results from HNC-EL/0 are shown as lines (small circles indicating the
equilibrium density $\rho_{\rm eq}$), the open symbols
are from corresponding PIMC simulations.  The filled symbol indicates
$E/N$ and $\rho_{\rm eq}$ estimated from a PIMC simulation of a
self-bound puddle of 50 dipoles in each layer, see also~\cite{supp}.
Bottom panel: equilibrium density $\rho_{\rm eq}$ versus $d$.
}
\label{FIG:E0}
\end{figure}

{\em Results.}
We calculated the ground state energy per particle, $E(\rho)/N$, as function
of total density $\rho=\rho_A+\rho_B$ for different layer distances $d$.
The interlayer DDI scales with $d^{-3}$, therefore the energy per particle, $E/N$, varies over
a wide range, as can be seen in the top panel of Fig.~\ref{FIG:E0} that shows $E(\rho)/N$ for four values of $d$.
A key result is that $E(\rho)/N$ has a minimum at a certain equilibrium density $\rho_{\rm eq}(d)$,
where the pressure $p$ vanishes: without
an externally applied pressure provided e.g. by a trap potential,
the total density of the bilayer system will adjust itself to $\rho_{\rm eq}(d)$.  Rather
than expanding like a gas, a dipolar bilayer system with antiparallel
polarization is a {\em self-bound liquid}.  Despite the
purely repulsive intralayer interaction, the partly attractive
interlayer interaction provides the ``glue'' that binds the system
to a liquid.  The phenomenon of an effective intralayer attraction, mediated by
particles in the other layer, is discussed in more detail below.

During the iterative numerical optimization, convergence
becomes very sensitive as the density $\rho$ or the distance $d$ between layers $A$ and $B$ is decreased,
until the HNC-EL/0 equations eventually fail to converge.
Past experience with HNC-EL/0 is that a numerical
instability usually has a physical reason.  Indeed, as we will show below,
there is a spinodal point where the homogeneous phase assumed in our calculation
becomes unstable against phase separation by nucleation of puddles.  Thus the
equation of state $E(\rho)/N$ for a homogeneous phase indeed ends at a critical
density.

Also shown in the top panel of Fig.~\ref{FIG:E0} are the energies obtained with PIMC simulations.
The temperature is set to $T=0.5$ ($T=0.25$ for the smallest $\rho$),
which is low enough that the thermal effect on $E/N$ is smaller than the
symbol size.  The open triangles are bulk simulations of $N_A=N_B=50$ dipoles with
periodic boundary conditions.  The HNC-EL/0 results are upper bounds
on $E/N$, consistent with a variational approach.  The overall dependence of
$E/N$ on $\rho$ and $d$ is reproduced quite well with the HNC-EL/0 method,
which is orders of magnitude faster than PIMC simulations.  The black triangle
shows the energy from a PIMC simulation of $N_A=N_B=50$ dipoles and layer separation
$d=0.1$ {\em without} periodic boundary conditions.  Due to the liquid
nature of the bilayer, the dipoles indeed coalesce into a puddle of finite density, given by
$\rho_{\rm eq}(d)$ apart from corrections due to the surface line tension.
The density corresponding to the filled triangle is obtained from the
radial density profile $\rho(r)$ at $r=0$ (see~\cite{supp}) where $r$ is defined
relative to the center of mass of the puddle.  Thermal evaporation was suppressed by
choosing a low temperature of $T={1/16}$.  Although this simulation of a finite
cluster is not equivalent to bulk PIMC or HNC-EL/0 calculations, the
central density of the puddle is close to $\rho_{\rm eq}(d)$ from HNC-EL/0.

The equilibrium density $\rho_{\rm eq}$ as function of $d$ is shown in the bottom panel of
Fig.~\ref{FIG:E0}. $\rho_{\rm eq}(d)$ decreases rapidly
with increasing $d$.  For smaller $d$, the decrease is approximately
$\rho_{\rm eq}\sim d^{-3}$ and for larger $d$ it is closer to $\rho_{\rm eq}\sim d^{-4}$.
Based purely on the interlayer DDI $v_{AB}(r)$, one would
expect a scaling of $\rho_{\rm eq}$ with the inverse square of $r_{\rm min}=2d$,
leading to a scaling $d^{-2}$.  The deviation from $d^{-2}$
is due to the kinetic energy.  Only for very small
$d$ (very deep $v_{AB}(r)$), this simple picture can be expected to be valid, and indeed the
$\rho_{\rm eq}$-curve becomes less steep for smaller $d$ in the double logarithmic representation
in Fig.~\ref{FIG:E0}.  In this small-$d$ regime of extremely strong
interlayer correlation the HNC-EL method would not be reliable anymore.

\begin{figure}[ht]
\begin{center}
\includegraphics*[width=0.65\textwidth]{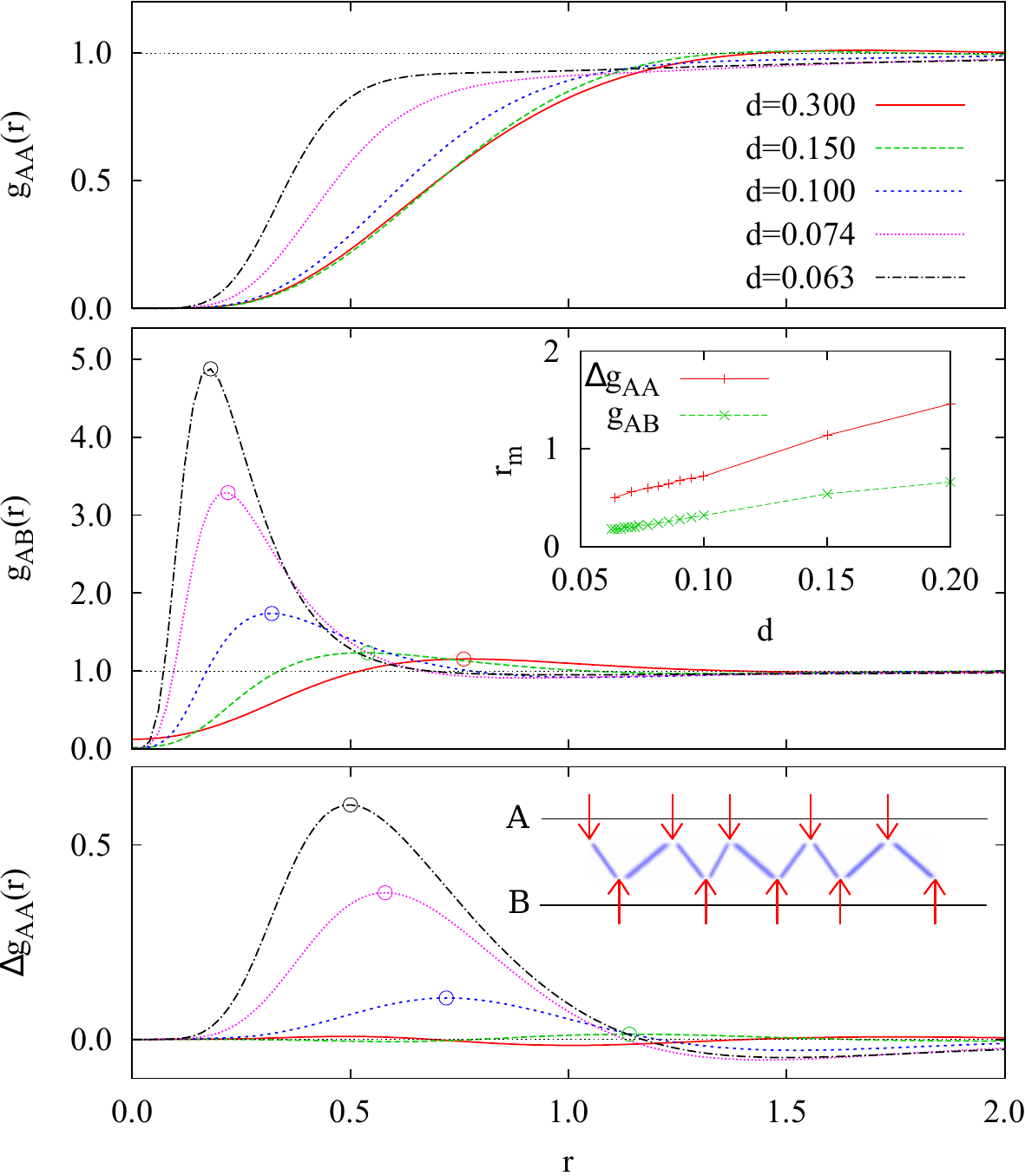}
\end{center}
\caption{(Color online)
Top panel: Intralayer pair distributions $g_{AA}(r)$ at density $\rho_A=\rho_B=0.5$
for progressively smaller layer distance $d$.
Middle panel: Corresponding interlayer pair distributions $g_{AB}(r)$,
with circle indicating the maxima of $g_{AB}(r)$.
Bottom panel: incremental intralayer pair distributions $\Delta g_{AA}(r)=g_{AA}(r)-g^\infty_{AA}(r)$,
i.e. the change from uncoupled layers.
The inset in the middle panel shows the positions of the maxima of
$g_{AA}(r)$ and $g_{AB}(r)$, respectively, as function of $d$.
The inset in the bottom panel sketches the attractive forces between dipoles in different layers.
}
\label{FIG:g}
\end{figure}

In Fig.\ref{FIG:g} we show the intralayer and interlayer pair distributions, $g_{AA}(r)$
and $g_{AB}(r)$, in the top and middle panel for progressively smaller layer
distance $d$ up to the smallest numerically stable value $d=0.063$, for 
$\rho=1$.  The growth of a strong correlation peak in $g_{AB}(r)$
as $d$ is decreased is a direct consequence of the increasingly
deep attractive well of $v_{AB}(r)$ around $r_{\rm min}=2d$. 
But also $g_{AA}(r)$ develops additional correlations, seen as a shoulder in the
top panel.  The additional correlations are best seen in the difference
to the uncoupled ($d=\infty$) limit $g^\infty_{AA}(r)$,
$\Delta g_{AA}(r)=g_{AA}(r)-g^\infty_{AA}(r)$, shown in the bottom panel.
This additional positive correlation between dipoles in the same layer
is mediated by dipoles in the other layer: the attraction
between a dipole in layer A and a dipole in layer B, that leads to a peak at
distance $r_{m}$ -- a bit larger than $2d$ due to zero-point motion --,
induces an effective attraction between the
dipole in A and another dipole in A, leading to peak at about twice the 
distance, $2r_{\rm m}$.  The inset in the middle panel shows the maxima of
$g_{AB}(r)$ and $\Delta g_{AA}(r)$ (indicated by circles in the plots of $g_{AB}(r)$ and $\Delta g_{AA}(r)$)
as function of distance $d$.
Indeed the peaks of $\Delta g_{AA}(r)$ are located at about twice the distance
of the peaks of $g_{AB}(r)$.
This effective intralayer attraction, induced by the interlayer attraction, is
illustrated by a simple 1D picture in the inset in the bottom panel, which also illustrates a preference for
a certain interparticle spacing, i.e.\ density, where ``dipole bridges''
(blue lines) can form.  The present 2D situation is more complicated, but our results for
the pair correlations demonstrate this picture is approximately valid.

\begin{figure}[h]
\begin{center}
\includegraphics*[width=0.64\textwidth]{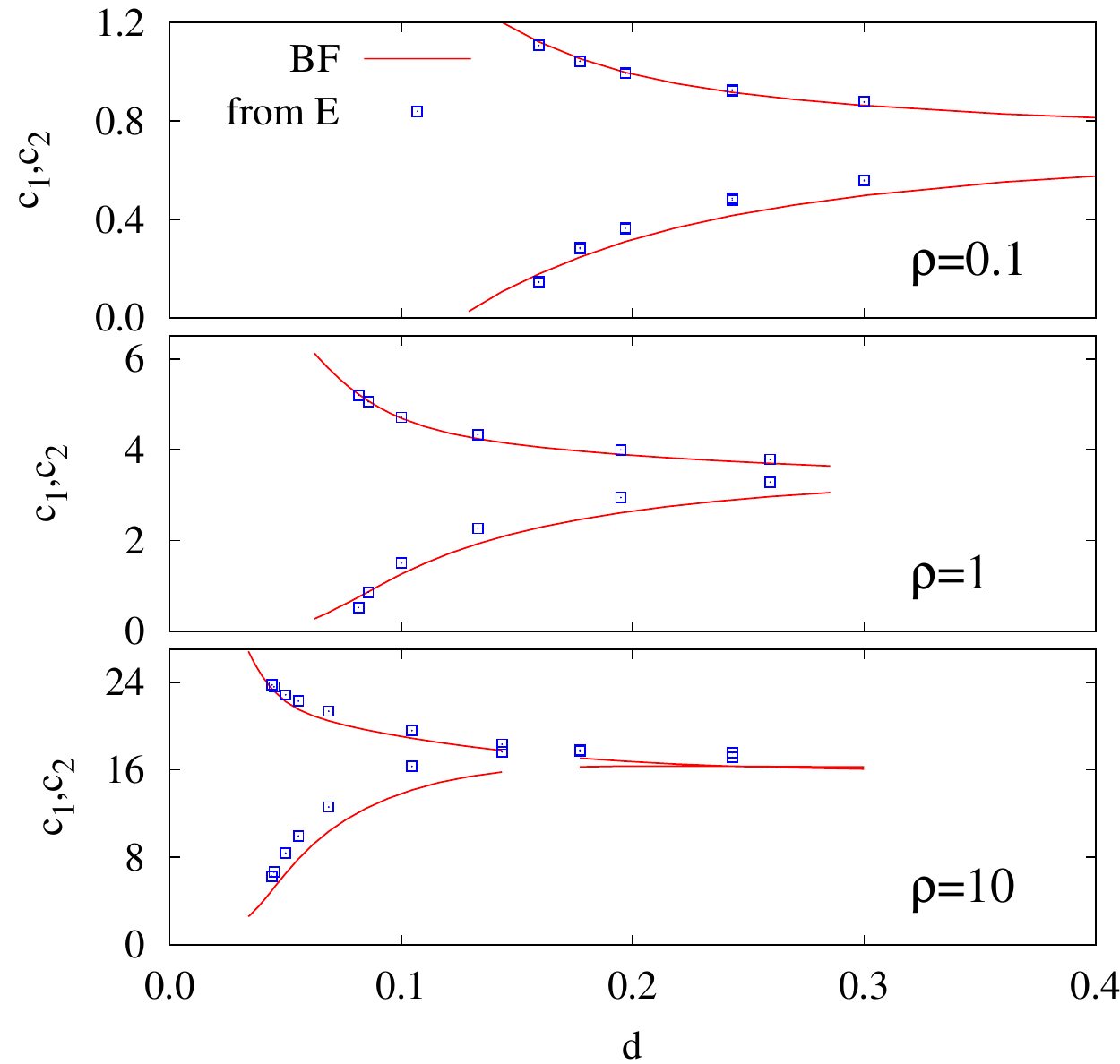}
\end{center}
\caption{
(Color online)
Speed of density fluctuations $c_1$ and speed of concentration fluctuations $c_2$, as function of
layer distance $d$ for three densities, $\rho=0.1;1;10$ (top, middle, and bottom panels).
Full lines show the \BFA\ and symbols are the thermodynamic estimates.
}
\label{FIG:c}
\end{figure}

The identification of the low density instability with a spinodal point can be proven
by calculating the long wavelength modes.  For two coupled layers, there are
two excitation modes, $\epsilon_{1,2}$ for any given wave number $k$, a density mode and a concentration
mode.  In the long wavelength limit, $k\to 0$, each mode can be characterized by
the speed of a density or concentration fluctuation, $c_1$ and $c_2$, respectively.
At the spinodal point, $c_1$ vanishes which means that the system
becomes unstable against infinitesimal $k\to 0$ fluctuations of the total density, triggering
the spinodal decomposition.  The easiest way to calculate $c_1$ and $c_2$ is the
\BFA\ (BFA) for the excitation energies $\epsilon_i(k)$, i.e.\ solving the
generalized eigenvalue problem ${k^2\over 2} \vec\phi = \varepsilon \qS(k) \vec\phi$.
For strong correlations, the BFA gives only a rough idea of the true excitation structure,
e.g.\ the BFA for the roton energy of superfluid $^4$He is off by a factor of two.  However,
it describes the low momentum limit of the dispersion relation very well, which is what we
need for $c_i$.  For a symmetric bilayer, the eigenvalues are
$\epsilon_{1,2}(k) = {k^2\over 2} (S_{AA}(k) \pm S_{AB}(k))^{-1}$
and the associated eigenvectors are $\vec\phi_{1}\sim (1,1)$ and $\vec\phi_{2}\sim (1,-1)$.
$\vec\phi_{1}$ describes total density fluctuations where particles
in different layers move in phase and $\vec\phi_{2}$ describes concentration
fluctuations, where particles in different layers move out of phase,
and the total density is constant.
For small $k$, $\epsilon_1(k)<\epsilon_2(k)$, i.e. the density mode has
lower energy than the concentration mode.  For $k\to 0$ we get
$c_{1,2}={1\over 2}(S_{AA}' \pm S_{AB}')^{-1}$, where we abbreviated the
derivatives $S'_{\alpha,\beta}=dS_{\alpha,\beta}(k)/dk|_{k=0}$.
For single-component Bose systems, it is known that
long-wave length limit of $S(k)$ obtained with HNC-EL are biased by the approximation
made for elementary diagrams (omitted here altogether).  This leads to an inconsistency
between the speed of sound $c$ obtained from the HNC-EL approximation for $S(k)$ and
the thermodynamic relation between $c$ and the energy,
$c^2={\partial\over\partial\rho}\rho^2{\partial\over\partial\rho} {E\over N}$ (in dipole units).
In order to assess the reliability of our results for $c_1$ and $c_2$,
we compare the BFA values obtained from $S_{\alpha,\beta}$,
$c_{1,2}={1\over 2}(S_{AA}' \pm S_{AB}')^{-1}$, with the generalization of the thermodynamic
relation between $c_{1,2}$ and the energy to binary systems,
$c_{1,2}^2 = \rho(e_{AA}\mp e_{AB})/2$ where $e_{\alpha\beta}$ is the second
derivative of $E/N$ with respect to $\rho_\alpha$ and $\rho_\beta$ \cite{campbellJLTP71}.

In Fig.~\ref{FIG:c} the results for $c_1$ and $c_2$ obtained with the two methods are shown as
function of layer distance $d$ for densities $\rho=0.1;1;10$.
As the coupling between layers is increased by reducing $d$, $c_1$ and $c_2$ behave differently.
The speed of concentration fluctuations $c_2$ increase (without
actually diverging) while the quantity of main interest, the speed of density fluctuations $c_1$
decreases to zero, in agreement with the interpretation of the instability
as a spinodal point.  The critical distance where $c_1$ vanishes
is lower for higher $\rho$, hence increasing the density for a given $d$ makes the system more stable.
The BFA for $c_{1,2}$ and their thermodynamic estimates agree qualitatively,
but differ especially for the interesting regime near the spinodal point where $c_1\to 0$.
The Bijl-Feynman values for $c_1$ appear to go to zero linearly and at slightly smaller $d$,
while the thermodynamic estimates approach zero more steeply, possible in a non-analytic fashion.
Unfortunately, these uncertainties
preclude a meaningful analysis of critical exponents for $c_1(d)$ or $c_1(\rho)$.
Monte Carlo simulations, including a finite size scaling analysis,
may shed more light on this question, but would certainly require very large simulations,
beyond the scope of this paper.

In conclusion, we have shown that a dipolar bilayer with antiparallel polarization
in the two layers constitute a self-bound liquid, evidenced by a minimum of
$E(\rho)/N$ at a finite density.
The bilayer relaxes to a stable equilibrium at a finite total density
and requires no external pressure coming from a trapping potential,
which makes is possible to study homogeneous quantum phases.
Comparison with exact PIMC simulations shows good agreement with results obtained
with the HNC-EL/0 method.  As expected for a liquid,
the equation of state ends at a spinodal point where the bilayer becomes unstable
against infinitesimal long-wavelength perturbations, thus the speed of
total density fluctuations approaches zero.
Finite size PIMC simulations confirm that the system indeed coalesces into
a puddle with a flat density profile given by the equilibrium
density~\cite{supp}.

\begin{acknowledgments}
We acknowledge financial support by the Austrian Science Fund FWF (grant No.\ 23535),
and discussions with Ferran Mazzanti, Jordi Boronat, and Gregory Astrakharchik.
\end{acknowledgments}



%

\newpage

\centerline{\large\bf Supplement}


{\bf Comparison with PIMC}\medskip

An important question is how accurate are our ground state results obtained with the variational
HNC-EL/0 method, where we neglect elementary diagrams and higher than two-body
correlations in the ansatz for the wave function.  We performed path integral Monte Carlo
(PIMC) simulations to assess the qualtity of our HNC-EL/0 results and found good
agreement between HNC-EL/0 energies and PIMC energies at low temperature.  In this supplement
we present additional comparisons of the static structure matrix, as well
as results for finite systems, where self-bound ``puddles'' are formed.
All simulations were done with 50 particles per layer; bulk simulations used
quadratic simulation boxes with periodic boundary conditions with a box size
adjusted to achieve a given density.  For antiparallel bilayers, cut-off
corrections to the dipole-dipole interaction cancel each other and therefore are not needed.

\begin{figure}[ht]
\begin{center}
\includegraphics*[width=0.6\textwidth]{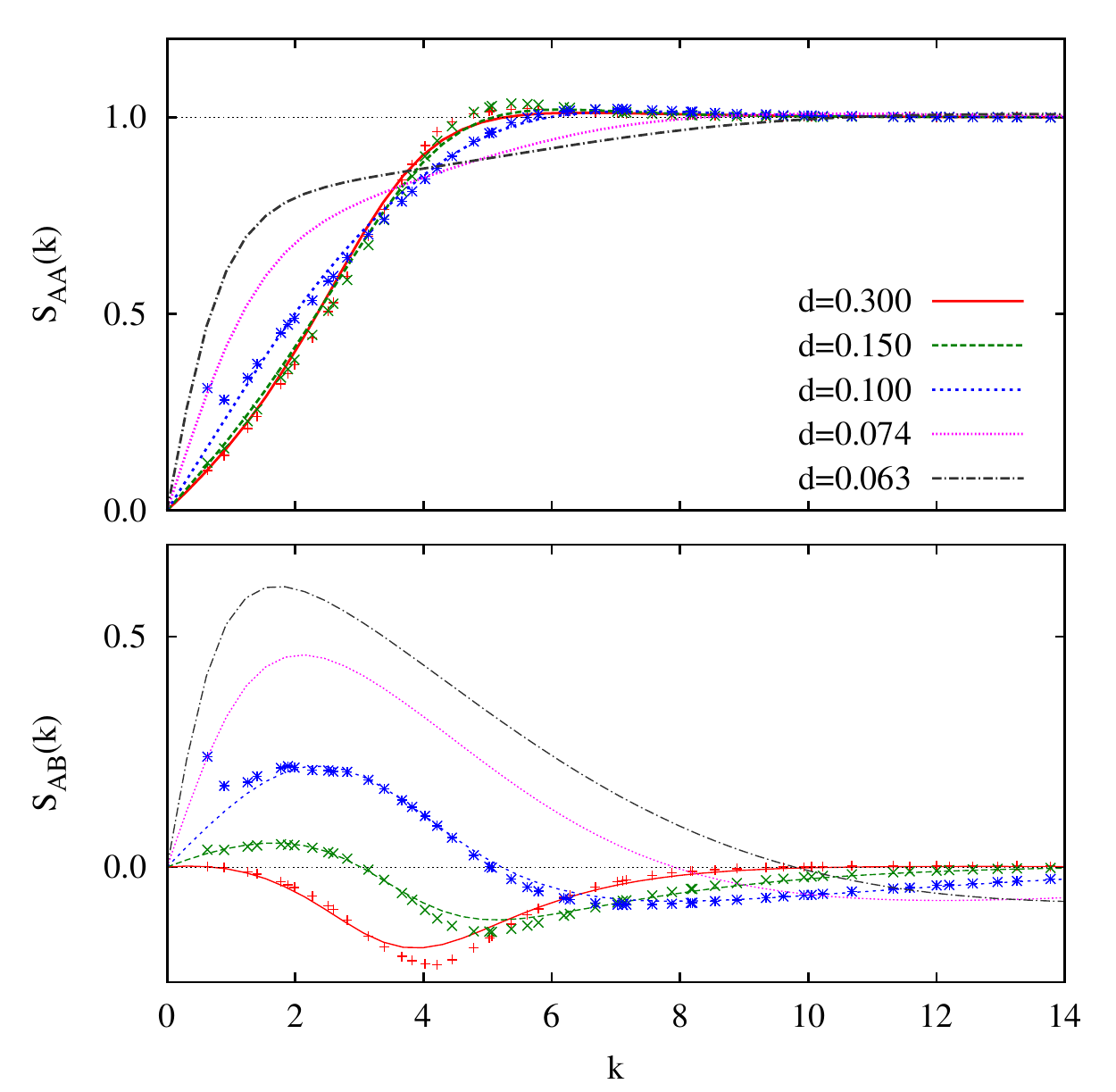}
\end{center}
\caption{
(Color online)
Static structure functions
$S_{AA}(k)$ (upper panel) and $S_{AB}(k)$ (lower panel) obtained with HNC-EL/0
(lines) for a total density $\rho=1$ and progressively smaller layer distances $d$
as indicated in the upper panel.  The symbols show $S_{AA}(k)$ and $S_{AB}(k)$ obtained
by PIMC simulations at $T=0.5$K for distances down to $d=0.1$.
}
\label{FIG:compare}
\end{figure}

\begin{figure}[ht]
\begin{center}
\includegraphics*[width=0.45\textwidth]{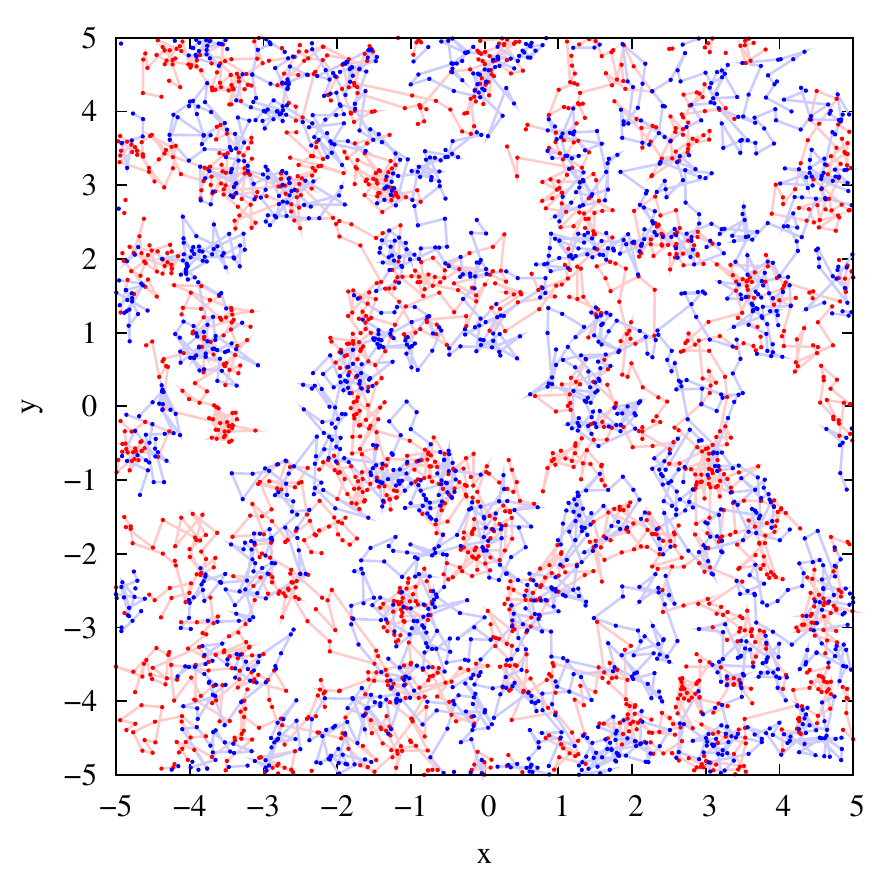}
\end{center}
\caption{
(Color online)
PIMC simulations snapshot for $\rho=1$ and $d=0.1$ ($T=0.5$).
Red and blue chains are the dipoles in layer A and B.
}
\label{FIG:snap}
\end{figure}

For predicting
the spinodal instability, but also for calculating excitation properties, an
important quantity is the static structure matrix, $S_{\alpha\beta}(k)$.  In Fig.\ref{FIG:compare}
we compare $S_{AA}(k)$ and $S_{AB}(k)$ obtained with HNC-EL/0 (lines) and PIMC (symbols),
at a total density of $\rho=1$.  The temperature in the PIMC simulation was $T=0.5$ which was
low enough that $S_{\alpha\beta}(k)$ did not change upon lowering the temperature.
We see that the HNC-EL/0 approximation works well, considering
that intralayer correlations are quite strong.  For $d=0.3$ and $d=0.15$ PIMC simulations
predicts slightly more pronounced peaks and troughs, but HNC-EL/0 calculations are faster by
several orders of magnitude.  For $d=0.1$, the agreement is
also very good, except for the two smallest $k$ values possible in a simulation
box of side length 10, $k=2\pi/10\approx 0.63$.
In the PIMC results, both $S_{AA}(k)$ and $S_{AB}(k)$ turn up sharply for this smallest $k$ value.
When we reduce $d$ even more, this apparent peak at $k=0$ grows very large.  This peak
has a very simple reason: as we approach the spinodal point by reducing $d$, we enter the metastable
regime of the phase diagram, where $E/N$ as function of total density
has a negative slope.  In this regime a finite perturbation can lead to a collapse and
the system phase separates.  Since there is no trial wave function in PIMC that could
prevent that, this collaps indeed happens as we go to far below the equilibrium density.
Indeed, Monte Carlo snapshots such as in Fig.~\ref{FIG:snap}
show density fluctuations already for $d=0.1$ that resemble small ``bubbles''.  In Fig.~\ref{FIG:snap}
red and blue dots, connected by lines, are the beads of the discretized imaginary
time paths sampled in PIMC; each bead is a particle at a discrete time step.
For even lower $d$ or lower total densities we observe a clear decomposition into a droplet and a low-density
gas, see next paragraph.  A large peak at $k=0$ can then be seen as a zero momentum Bragg peak
due to phase separation.

\begin{figure}[ht]
\begin{center}
\includegraphics*[width=0.6\textwidth]{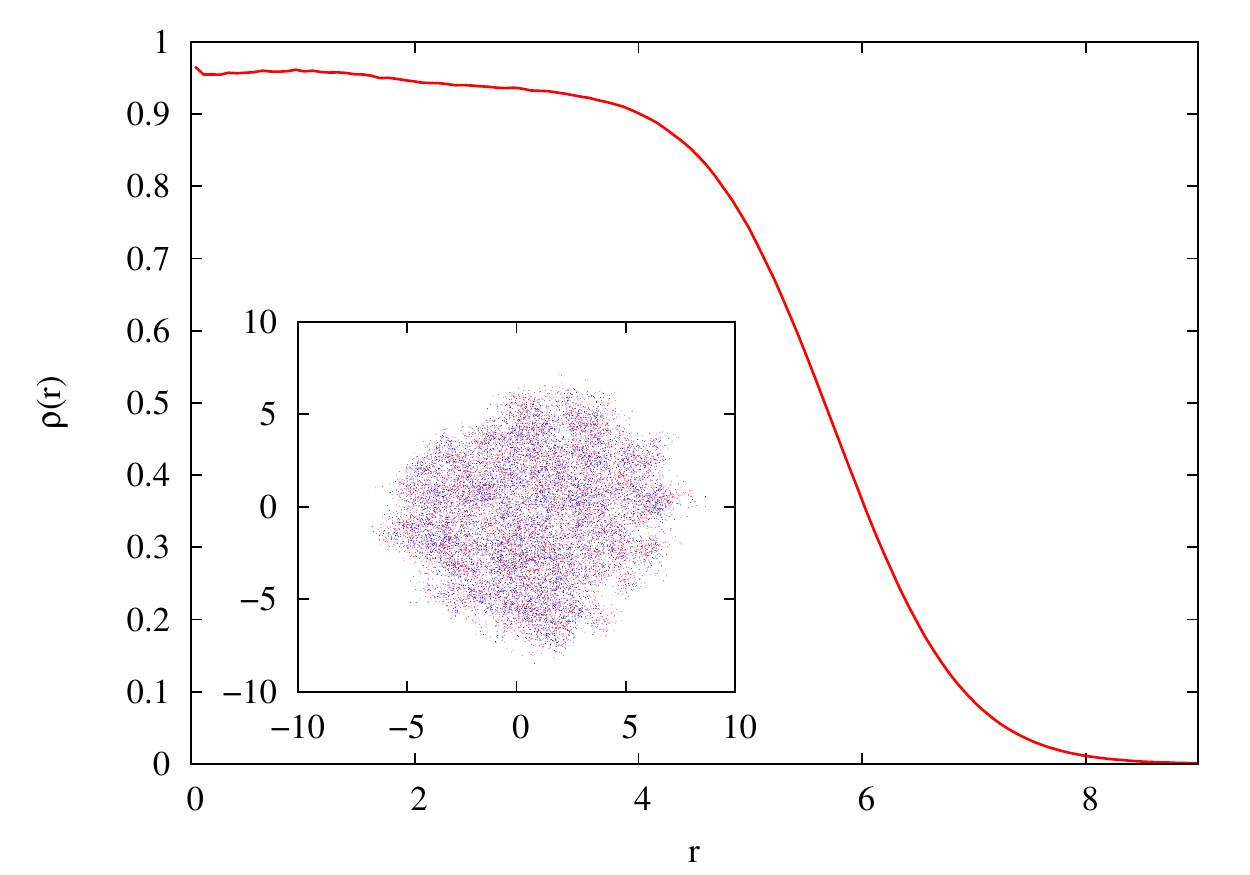}
\end{center}
\caption{
(Color online)
Density profile $\rho(r)$ for a self-bound droplet of 50 dipoles in each layer.
Inside the droplet the density is approximately constant, with a value close to
the equilibrium density $\rho_{\rm eq}$ of a bulk system at zero pressure.
The distance is $d=0.1$ and the temperature was set to $T={1\over 16}$, which
is low enough to prevent evaporation.
The inset shows a snapshot of the simulation, with red and blue indicating
the dipoles in the layers A and B, respectively, at the imaginary time steps of
the paths of PIMC.
}
\label{FIG:densprofile}
\end{figure}

As final confirmation of the liquid nature of dipolar bilayers with antiparallel
polarization, we show the results of a PIMC simulation without periodic boundary
conditions and without any external trap potential in the planar direction.
A two-dimensional gas would of course spread out indefinitely, while a liquid
will coalesce into a droplet of finite density.  For a large enough droplet such
that effects of surface line tension are negligible, the density inside the droplet
is given by the equilibrium density $\rho_{\rm eq}$, i.e. the density of a bulk system
at zero pressure.  In Fig.\ref{FIG:densprofile} we show the radial density profile
$\rho(r)$ for 50 dipoles in each layer separated by $d=0.1$,
where $r$ is measured relative to the center of mass.  In order to prevent evaporation we
set the temperature to $T={1\over 16}$.  $\rho(r)$ is approximately constant
for $r\lesssim 4$ and quickly falls to zero for larger $r$.  This is the
behavior expected for the density profile of a self-bound liquid, and very different from the density
profile of a quantum gas in a trap.

\end{document}